\documentclass{IEEEtran}
\usepackage{graphicx}
\usepackage{amsmath}
\usepackage{multicol}
\usepackage{multirow}
\usepackage{stfloats}
\usepackage{booktabs}
\usepackage{mathrsfs}
\usepackage{enumerate}
\usepackage{amssymb}
\usepackage{algorithm}
\usepackage{array}
\usepackage{underscore}
\usepackage{url}
\usepackage{hyperref}
\hypersetup{hidelinks}
\usepackage{flushend}

\usepackage{algorithmic}
\usepackage{graphicx}
\usepackage{epstopdf}
\usepackage{caption}
\usepackage{diagbox}

\usepackage{caption}

\usepackage{subfigure}
\usepackage{color}

\usepackage[outerbars,color]{changebar}
\ifx\pdfoutput\undefined
\else\ifnum\pdfoutput>0
  \usepackage{pdfcolmk}
\fi\fi
\cbcolor{red}
\usepackage{fancyhdr}

\makeatletter
\newcommand{\Rmnum}[1]{\expandafter\@slowromancap\romannumeral #1@}
\makeatother

\usepackage{array}  \newcommand{\PreserveBackslash}[1]{\let\temp=\\#1\let\\=\temp}  \newcolumntype{C}[1]{>{\PreserveBackslash\centering}p{#1}}  \newcolumntype{R}[1]{>{\PreserveBackslash\raggedleft}p{#1}}  \newcolumntype{L}[1]{>{\PreserveBackslash\raggedright}p{#1}}

\ifCLASSINFOpdf
\else
\fi
\begin{document}
\title{\begin{huge}Optimal Resonant Beam Charging for Electronic Vehicles\\ in Internet of Intelligent Vehicles\end{huge}}

\author{Qingqing~Zhang,
Mingqing~Liu,
Xing~Lin,
Qingwen~Liu\IEEEauthorrefmark{1},
Jun~Wu, 
and~Pengfei~Xia	

	\thanks{Q.~Zhang, M. Liu, X. Lin, Q. Liu, J. Wu, and P. Xia, are with  College of Electronics and Information Engineering, Tongji University, Shanghai, People's Republic of China, e-mail: anne@tongji.edu.cn, clare@tongji.edu.cn, linxing018@tongji.edu.cn, qliu@tongji.edu.cn,  wujun@tongji.edu.cn, and pengfei.xia@gmail.com.}%

\thanks{* Corresponding author.}

\thanks{Copyright (c) 2012 IEEE. Personal use of this material is permitted. However, permission to use this material for any other purposes must be obtained from the IEEE by sending a request to pubs-permissions@ieee.org.}
}

\maketitle

\begin{abstract}
To enable electric vehicles (EVs) to access to the internet of intelligent vehicles (IoIV), charging EVs wirelessly anytime and anywhere becomes an urgent need. The resonant beam charging (RBC) technology can provide high-power and long-range wireless energy for EVs. However, the RBC system is unefficient. To improve the RBC power transmission efficiency, the adaptive resonant beam charging (ARBC) technology was introduced. In this paper, after analyzing the modular model of the ARBC system, we obtain the closed-form formula of the end-to-end power transmission efficiency. Then, we prove that the optimal power transmission efficiency uniquely exists. Moreover, we analyze the relationships among the optimal power transmission efficiency, the source power, the output power, and the beam transmission efficiency, which provide the guidelines for the optimal ARBC system design and implementation. Hence, perpetual energy can be supplied to EVs in IoIV virtually.
\end{abstract}

\begin{IEEEkeywords}
Internet of Things, Internet of Intelligent Vehicles, Resonant Beam Charging.
\end{IEEEkeywords}

\IEEEpeerreviewmaketitle

\section{Introduction}\label{Section1}
The fast growing mobile computing and communication applications of electric vehicles (EVs), such as electric cars, unmanned aerial vehicles (UAVs) and electromobiles, accelerate the development of the internet of intelligent vehicles (IoIV) \cite{UAV,IoT,zhang2018flexible,zeng2017channel,cheng20175g,cheng2015d2d}. To enable the EVs to access to the IoIV anytime and anywhere, the batteries of EVs should be able to support their operations all the time \cite{cheng2014electrified,cheng2016consumer,zhang2016energy}. However, refilling the EVs' batteries faces the challenges of battery capacity limitation and power supply availability. Wired charging is inconvenient, because users have to seek for a power output and wait a long time for charging. Therefore, wireless charging or wireless power transfer (WPT) attracts great attention to provide perpetual energy supplies for EVs virtually \cite{zhang2018flexible,cheng2016consumer,87456,Feng2017Locator}.

The existing wireless charging technologies include: inductive coupling, magnetic resonance coupling, electromagnetic (EM) radiation, and resonant beam charging (RBC) \cite{wirelesstechniques,liu2016dlc,dlcqing}. In the RBC system, multiple devices can be charged simultaneously by one power transmitter, which is like Wi-Fi communications. Moreover, RBC can charge electrical devices with high power over a long distance \cite{liu2016dlc,wicharge,22222}. Therefore, RBC is suitable for charging EVs. To improve RBC efficiency, the adaptive resonant beam charging (ARBC) was presented in \cite{Qing2017,arbcqing}. The ARBC transmitter can adjust the source power according to the receiver preferred charging power based on feedback control. For charging a 1000mAh Li-ion battery, the ARBC system can save at least 60.4\% of charging energy compared with the RBC system \cite{Qing2017}.

\begin{figure}
	\centering
	\includegraphics[scale=0.25]{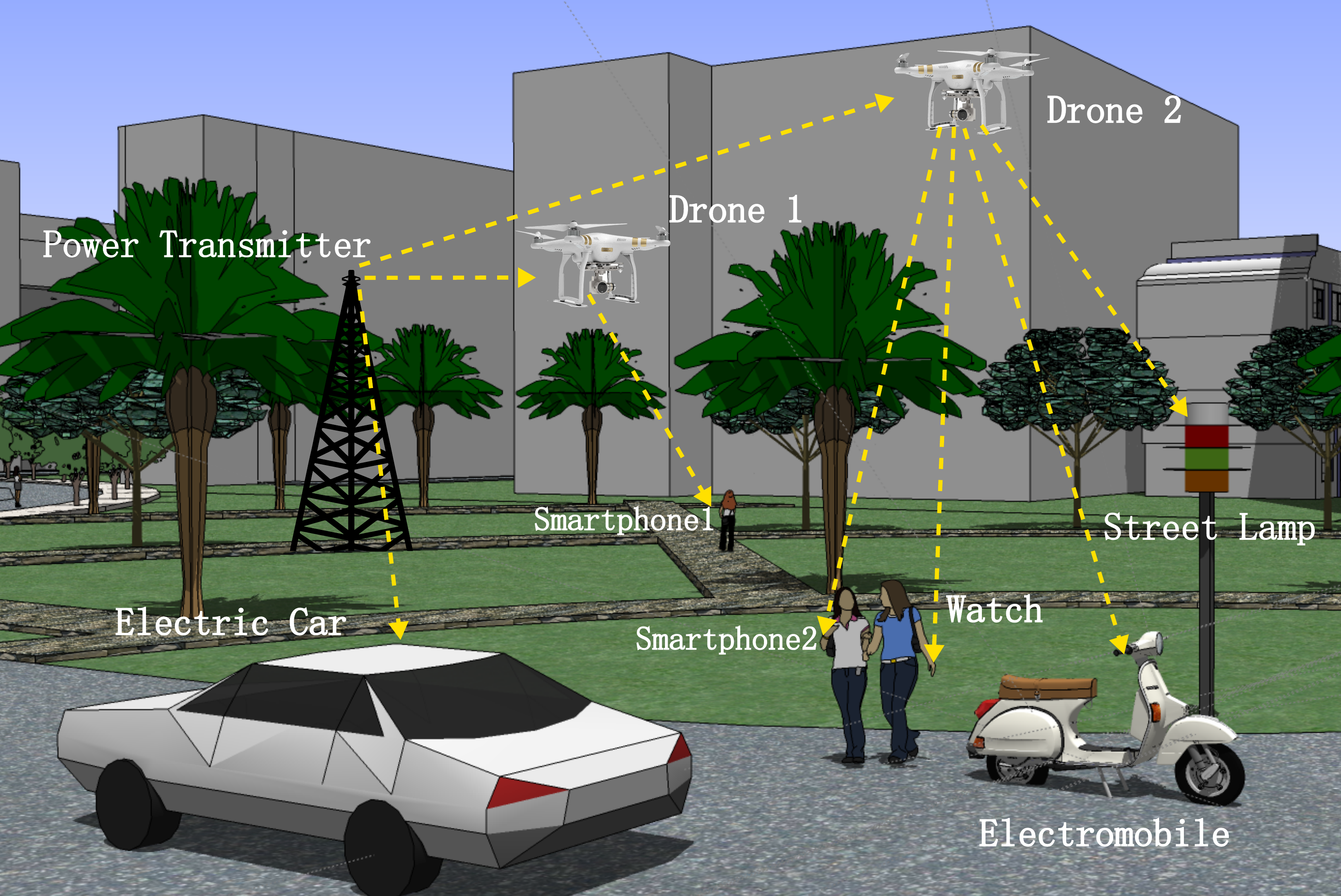}
	\caption{Adaptive Resonant Beam Charging Applications}
	\label{adhocnetwork}
\end{figure}

Fig.~\ref{adhocnetwork} gives an ARBC application scenario. An ARBC transmitter, which is like a base station in wireless communications, is installed to provide wireless power to devices in its coverage. The electric car, drone 1, and drone 2 are all in the transmitter's coverage, so they can be charged wirelessly. At the same time, the two drones can charge devices within their own coverage, i.e., work as relays. Drone 1 can charge smartphone 1, while drone 2 supports charging smartphone 2, watch, electromobile and street lamp simultaneously.

\begin{figure}[b]
	\centering
	\includegraphics[scale=0.4]{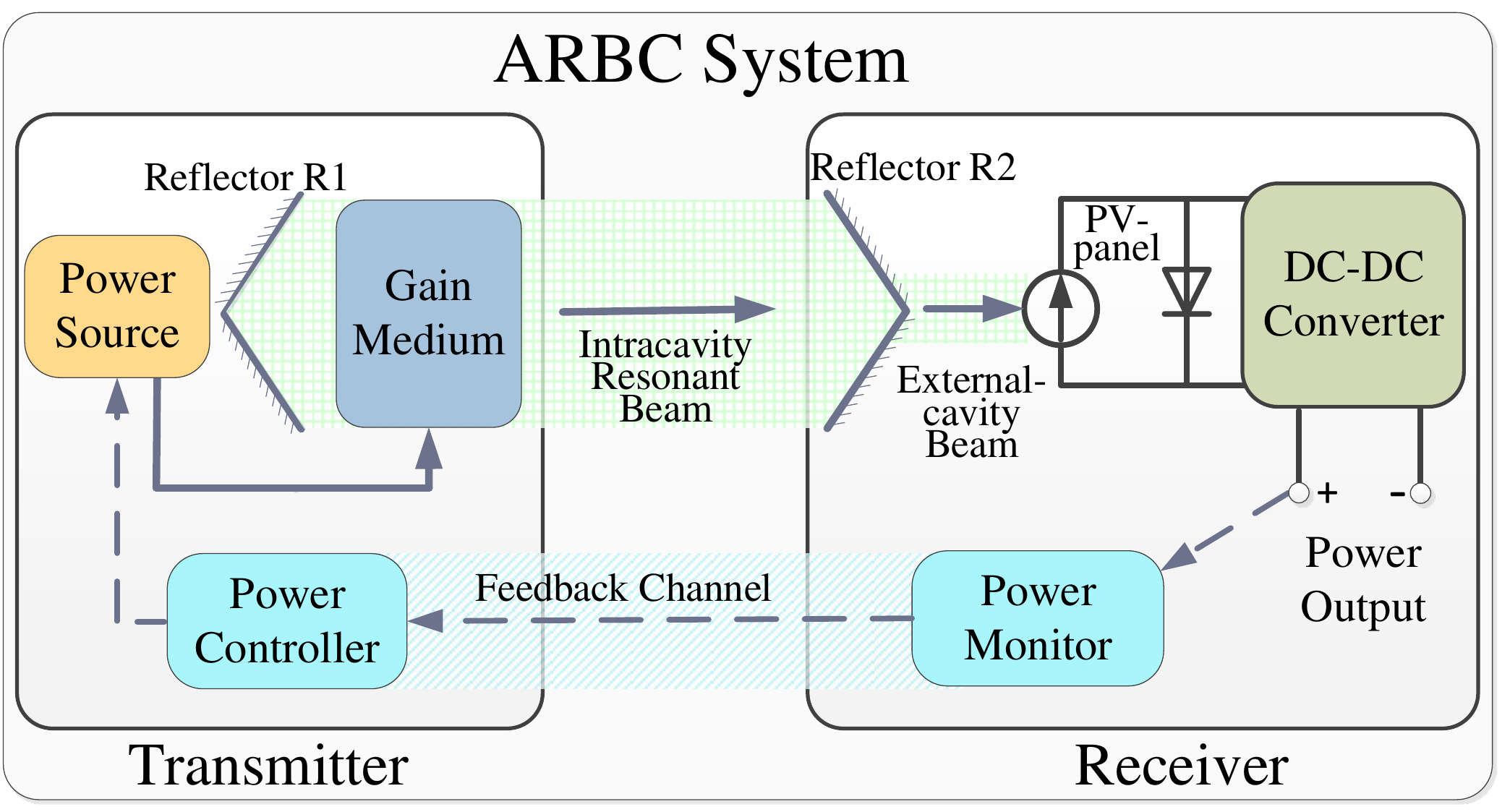}
	\caption{Adaptive Resonant Beam Charging System Diagram}
	\label{adaptivepower}
\end{figure}

To demonstrate the mechanism of the ARBC system, Fig.~\ref{adaptivepower} shows the ARBC system diagram \cite{Qing2017}. The ARBC system is separated into multiple conceptually independent modules: 1) the electricity-to-beam conversion module: the power source provides electrical power to stimulate the gain medium under the control of the power controller, and then the intra-cavity resonant beam power can be generated. 2) the beam transmission module: the intra-cavity resonant beam travels through the air and arrives at the ARBC receiver with attenuation. 3) the beam-to-electricity conversion module: the intra-cavity resonant beam is partially transformed to the external-cavity beam, which can be converted to the electrical power by a photovoltaic-panel (PV-panel). The direct current-to-direct current (DC-DC) converter is adopted to convert the PV-panel output current and voltage to the preferred output values. 4) the feedback module: the power monitor gets the preferred output current and voltage, thus power, and sends the values back to the power controller through the feedback channel. Repeating this procedure, electric vehicles can be charged with the device preferred current and voltage.

The contributions of this paper include: 1) We prove that the optimal power transmission efficiency uniquely exists, based on the closed-form formula of the end-to-end maximum power transmission efficiency of the ARBC system; 2) We analyze the relationships among the optimal power transmission efficiency, the source power, the beam transmission efficiency, and the output power, which provide the guidelines for the optimal ARBC system design and implementation.

In the rest of this paper, we will at first discuss the ARBC system modeling. Then, we will analyze the end-to-end performance of the ARBC system. Finally, we will make a conclusion and discuss the open issues.


\section{System Modeling}\label{Section2}
In this section, we will describe the mathematical models of the ARBC system, including the electricity-to-beam conversion, the beam transmission and the beam-to-electricity conversion.

\subsection{Electricity-to-Beam Conversion}\label{}
\begin{figure}
	\centering
    \includegraphics[scale=0.6]{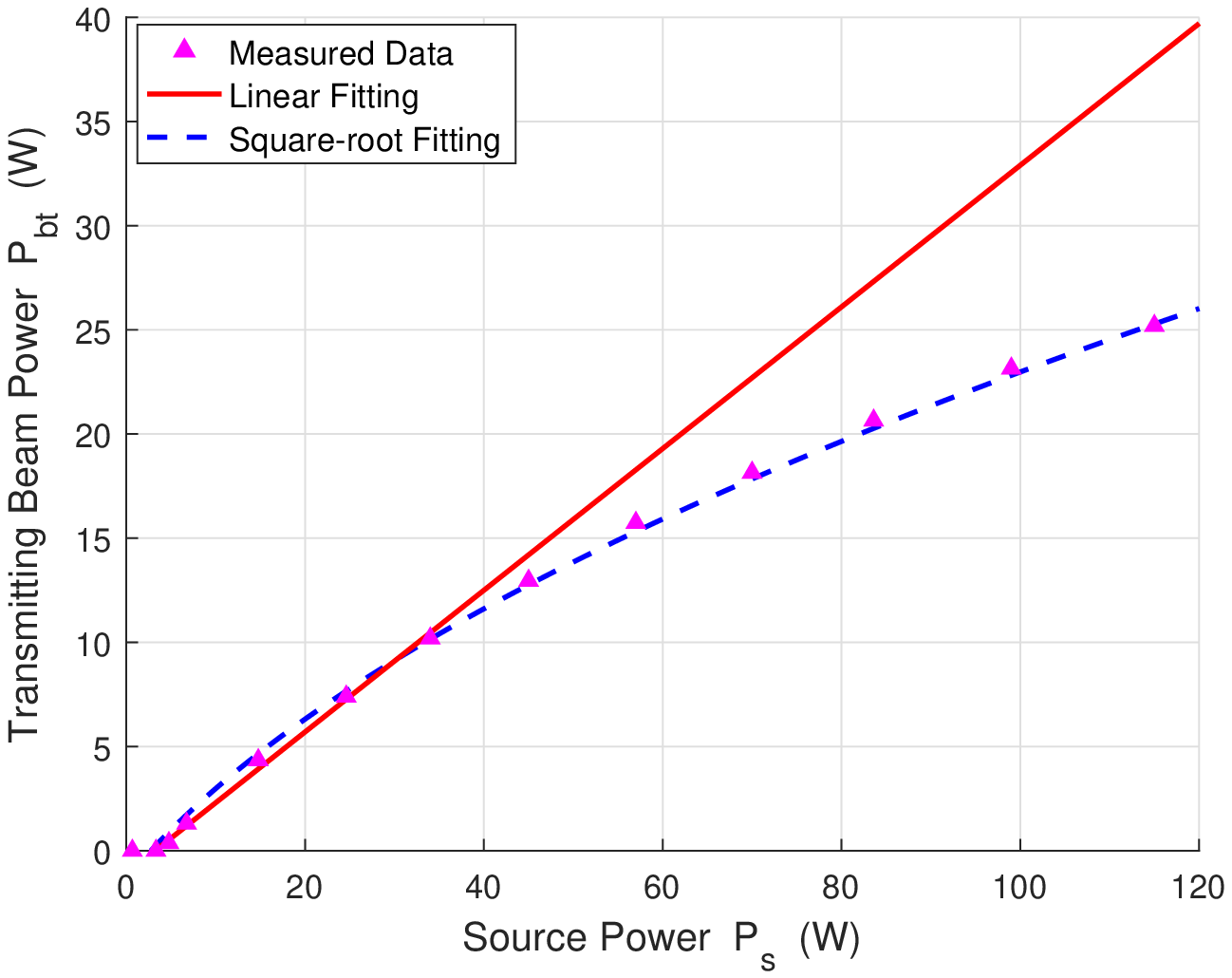}
	\caption{Transmitting Beam Power vs. Source Power}
    \label{1550plps}
%
\vspace{12pt}
	\centering
    \includegraphics[scale=0.6]{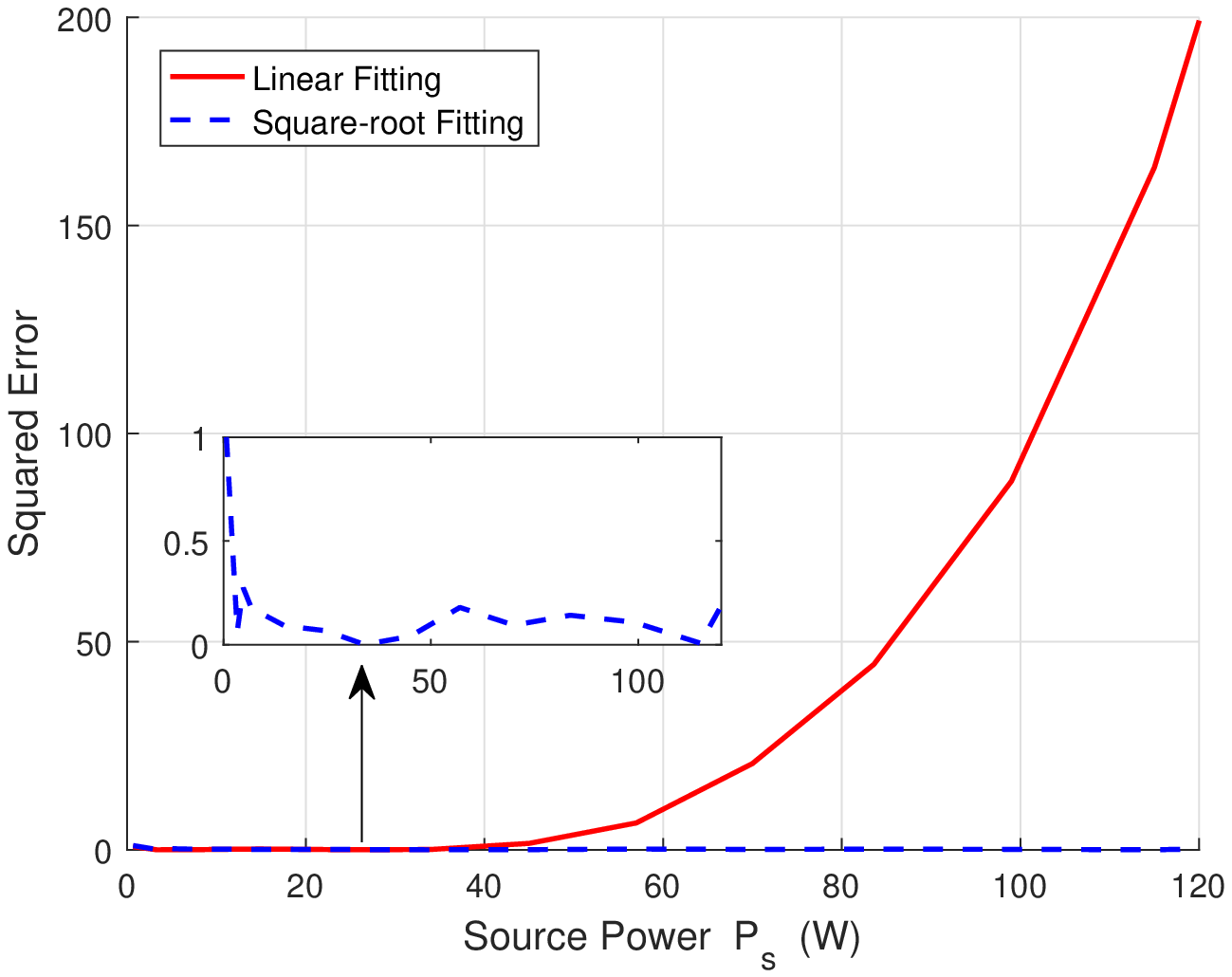}
	\caption{Squared Error vs. Source Power}
    \label{1550sse}
\end{figure}

At the ARBC transmitter, the power controller informs the power source to generate the corresponding electrical source power $P_s$, which can stimulate the gain medium to generate the intra-cavity resonant beam power at the transmitter $P_{bt}$. From \cite{Qing2017}, $P_{bt}$ varies with different beam wavelengths. When the distance between the transmitter and the receiver is 0, the measured values of $P_s$ and $P_{bt}$ for the 1540-1560nm beam system can be obtained from \cite{1550nmtransmitter}. Therefore, the relationship between $P_s$ and $P_{bt}$ can be described. The triangles in Fig.~\ref{1550plps} show the measured data points for 1550nm. As can be seen, the beam power can be stimulated out only when the source power is over a certain threshold since extra power consumption caused by other factors, such as thermal effect, is inevitable. 

To depict the relationship between $P_{bt}$ and $P_{s}$ when $P_{s}$ is over the certain threshold quantitatively, we use the following square-root fitting method:
\begin{equation}\label{plps}
P_{bt}\approx a_1 \sqrt{b_1+P_{s}} + c_1,
\end{equation}
where $a_1$, $b_1$, and $c_1$ are three coefficients, which are listed in Table \ref{ebparamaters}.

In Fig.~\ref{1550plps}, the solid-line gives the linear curve-fitting relationship between $P_{bt}$ and $P_{s}$, which is detailed in \cite{Qing2017}. The dash-line shows the square-root  curve-fitting relationship between $P_{bt}$ and $P_{s}$. As can be seen, the dash-line matches the triangles very well, while the solid-line can not match the triangles when $P_{s}$ is over 40W. This illustrates that the square-root approximation in \eqref{plps} is more consistent with the measurements than the linear one.

\begin{table}[bp]
\newcommand{\tabincell}[2]{\begin{tabular}{@{}#1@{}}#2\end{tabular}}
\centering
\caption{Electricity-to-Beam Conversion Parameters}
\begin{tabular}{C{1.5cm} C{1.5cm}}
\hline
\textbf{Symbol} & \tabincell{c}{\textbf{Value}} \\
\hline
\bfseries{$a_1$} & \tabincell{c}{$3.331$} \\
\bfseries{$b_1$} & \tabincell{c}{$10.2$} \\
\bfseries{$c_1$} & \tabincell{c}{$-11.99$} \\
\hline
\label{ebparamaters}
\end{tabular}
\end{table}

To evaluate the two fitting methods, we take the squared error based on the measured data in \cite{1550nmtransmitter} to depict the relative deviation between the fitting values and the measured values. The squared error $S_{e}$ is calculated as:
\begin{equation}\label{se}
\begin{aligned}
S_{e} = (P_{bt}-P_{s})^2,
\end{aligned}
\end{equation}
where $P_{bt}$ can be the linear-fitting beam power value or the square-root fitting beam power value. Fig.~\ref{1550sse} shows the errors of the linear-fitting method and the square-root fitting method. It can be seen that squared error of the linear-fitting method is larger than that of the square-root fitting method.

Based on the squared error, the mean square error (MSE) of the two methods can be obtained according to:
\begin{equation}\label{mse}
\begin{aligned}
MSE = \frac{1}{n}\sum_{i=1}^n [P_{bt(i)}-P_{s(i)}]^2,
\end{aligned}
\end{equation}
where $n$ is the number of the measured data. The MSE of the linear-fitting method is 43.5967, while that of the square-root fitting method is only 0.2057. Therefore, the measured data can be better explained by the square-root fitting method rather than the linear-fitting method.

\begin{figure}
	\centering
    \includegraphics[scale=0.6]{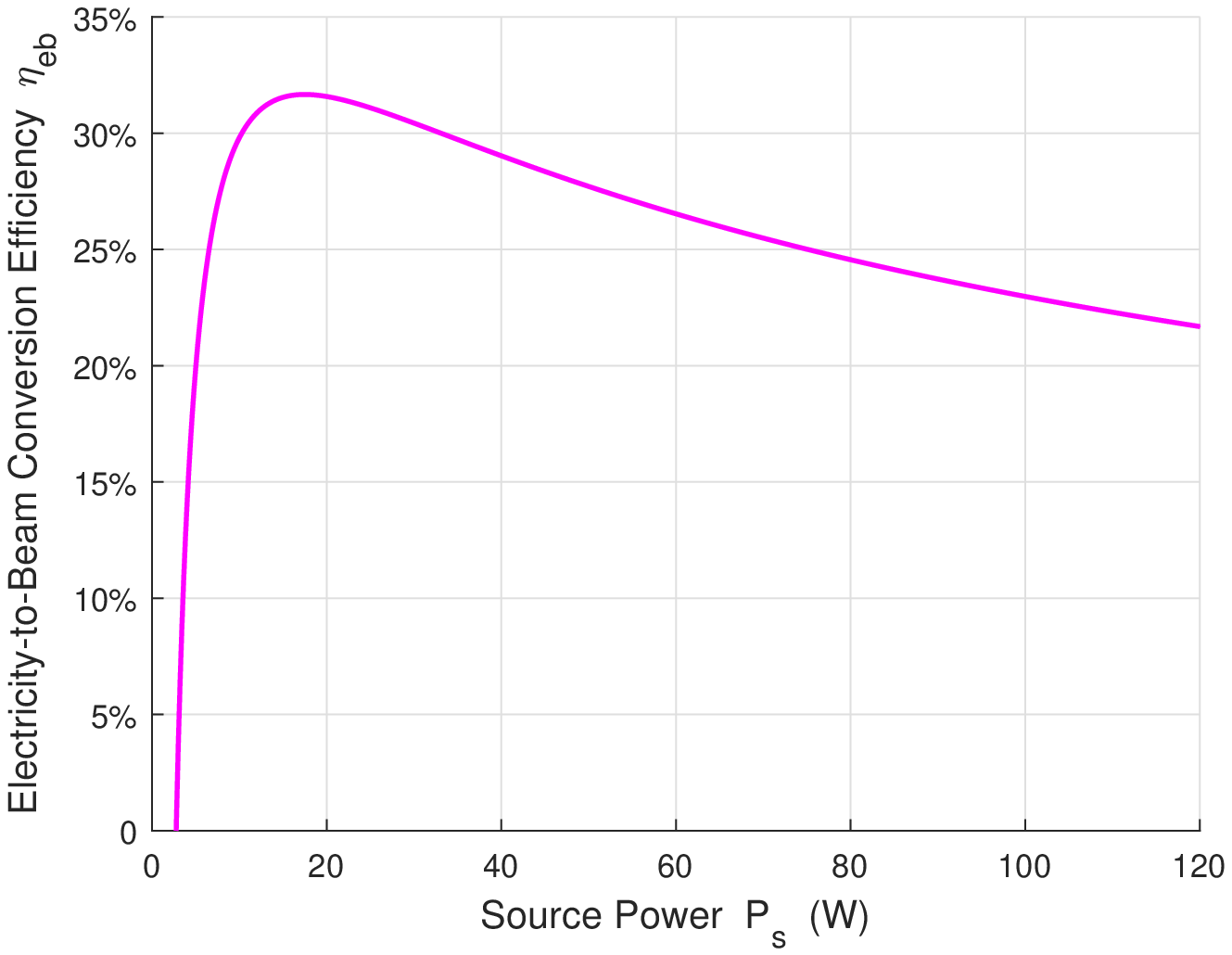}
	\caption{Electricity-to-Beam Conversion Efficiency vs. Source Power}
    \label{1550etaelf}
	\centering
    \includegraphics[scale=0.6]{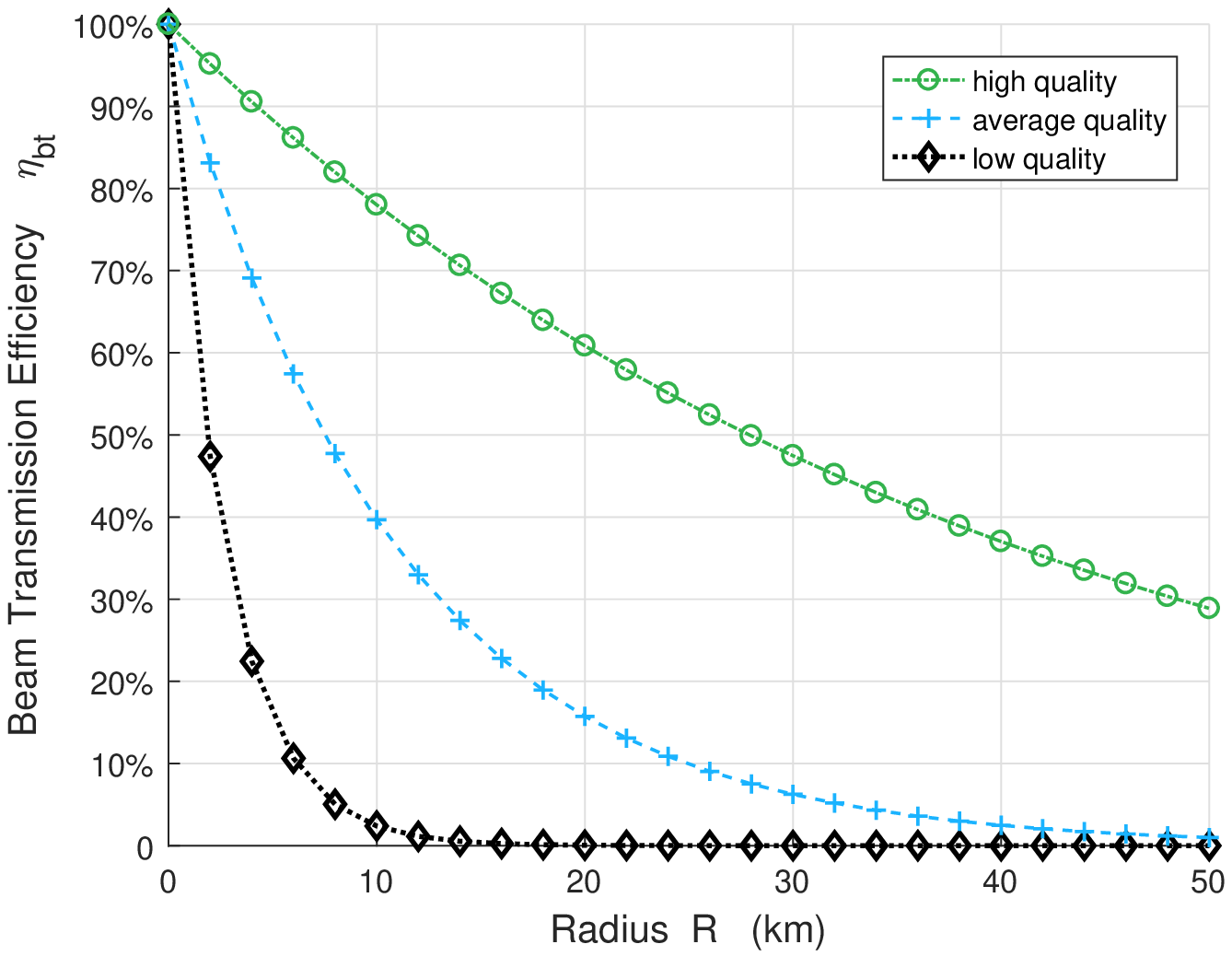}
	\caption{Beam Transmittance Efficiency vs. Transmission Radius}
    \label{transmittance}
\end{figure}

Based on \eqref{plps}, the electricity-to-beam conversion efficiency $\eta_{eb}$ can be obtained as:
\begin{equation}\label{etaeb}
  \eta_{eb} = \frac{P_{bt}}{P_s} \approx \frac{a_1 \sqrt{b_1+P_{s}} + c_1}{P_s}.
\end{equation}

From \eqref{etaeb}, $\eta_{eb}$ relies on $P_{s}$. Fig.~\ref{1550etaelf} illustrates how $\eta_{eb}$ changes with $P_{s}$. When the source power $P_{s}$ is over the threshold, $\eta_{eb}$ starts to increase dramatically until reaching the peak. The peak value is about 31.7\%. Then, $\eta_{eb}$ tends to decline against the peak slightly as $P_{s}$ increases.

\subsection{Beam Transmission}\label{}
After being stimulated out by the electrical power at the transmitter, the intra-cavity resonant beam transmits through the air to the receiver. During the transmission procedure, the beam power suffers from attenuation, similar to EM wave propagation power loss \cite{beampropagation}. 
Without attenuation, the external beam power at the receiver $P_{br}$ is equal to the intra-cavity resonant beam power at the transmitter $P_{bt}$.

The factors affecting the beam power transmission include the transmission range and the air quality \cite{Qing2017, karatay2004alternative}. To quantify the attenuation, we assume that the beam diameter is a constant. This assumption could be validated by controlling aperture diameters of the transmitter and the receiver.

The beam transmission efficiency $\eta_{bt}$ is modeled as \cite{karatay2004alternative}:
\begin{equation}\label{etabt}
  \eta_{bt}=\frac{P_{br}}{P_{bt}}= e^{-\frac{3.91}{\nu} \Big(\frac{\lambda}{550nm}\Big)^{-\chi} R},
\end{equation}
where $\nu$ is the atmospheric visibility, $\lambda$ is the beam wavelength in nm, $\chi$ is the size distribution of the scattering particles, and $R$ is the radius of the transmission range. For different air quality, $\nu$ and $\chi$ take different values. Here we take three typical scenarios, i.e., high visibility, average visibility, and low visibility, into consideration. For the three scenarios, $\chi$ can be specified as:
\begin{equation}\label{chic}
\chi = \left\{
             \begin{array}{lr}
             1.6,  \qquad\quad\  \mathrm{high \ visibility} \qquad\  (21km\leq \nu \leq 50km)&  \\
             1.3,  \qquad\quad\ \mathrm{average \ visibility} \quad (6km\leq \nu \leq 21km)\\
             0.585\nu^\frac{1}{3}. \quad\  \mathrm{low \ visibility} \qquad\ \  (\nu \leq 6km) &
             \end{array}
\right.
\end{equation}

Fig.~\ref{transmittance} depicts the variation of $\eta_{bt}$ with $R$ under the three different scenarios when $\lambda$ takes 1550nm. The parameters are given in Table \ref{pathloss}. It is clear that $\eta_{bt}$ decreases exponentially with the increment of $R$. The beam power attenuation depends on the visibility $\nu$, which is determined by the air quality. With the same transmission radius, high visibility brings high transmission efficiency.

\subsection{Beam-to-Electricity Conversion}\label{}

\begin{figure}
	\centering
    \includegraphics[scale=0.6]{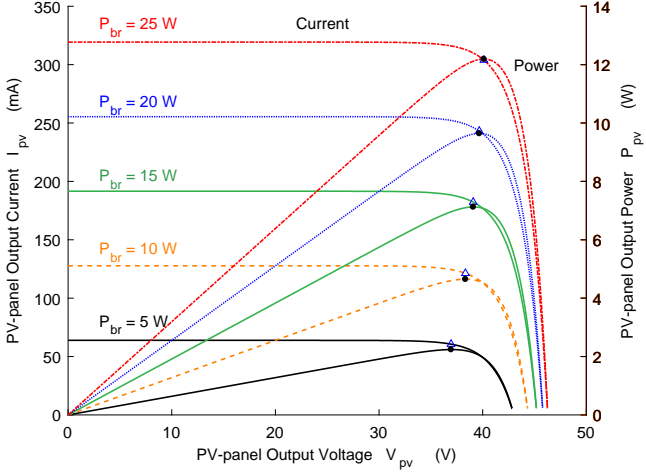}
	\caption{PV-panel Output Current and Power vs. Voltage ($T$=25$^{\circ}$C)}
    \label{1550irradiance}
\end{figure}

At the ARBC receiver, the PV-panel is adopted to transfer the external-cavity beam power to the electrical power. Factors influencing the beam-to-electricity conversion include the received beam power, beam wavelength, and PV-cell temperature \cite{Qing2017,Aziz2015Simulation}. In this section, we explore how these factors influence the beam-to-electricity conversion. Results here are obtained by using the standard \emph{solar cell} Simulink model \cite{solarcell}.

Fig.~\ref{1550irradiance} demonstrates the relationships among the PV-panel output current $I_{pv}$, power $P_{pv}$ and voltage $V_{pv}$ under different receiver beam power $P_{br}$ at room temperature (25$^{\circ}$C) for the GaSb-based PV-panel with 1550nm beam. All the parameters are listed in Table \ref{pvparamaters}.

From Fig.~\ref{1550irradiance}, given $P_{br}$, $I_{pv}$ keeps almost a constant when $V_{pv}$ is below a turning point. However, when $V_{pv}$ is over the turning point, $I_{pv}$ drops rapidly. For the same $V_{pv}$, $I_{pv}$ increases as $P_{br}$ decreases. When $I_{pv}$ is close to zero, $V_{pv}$ is the open-circuit voltage, which increases as $P_{br}$ increases. Moreover, given $P_{br}$, $P_{pv}$ increases as $V_{pv}$ increases until it reaches the peak point, which is corresponding to the turning point of $I_{pv}$. However, $P_{pv}$ drops dramatically when $V_{pv}$ is above the corresponding voltage for the peak point. For a given voltage $V_{pv}$, the output power $P_{pv}$ increases when the input beam power $P_{br}$ increases.

As to the turning point in Fig.~\ref{1550irradiance}, it is the maximum point of the output power, which is defined as the maximum power point (MPP). From \cite{onlyMPP2}, given the certain scenario and $P_{br}$, MPP is proved to be an unique point. For maximum efficiency, the PV-panel is supposed to work at the MPP, which can be tracked with the maximum power point tracking (MPPT) technology \cite{mppt}. We define $P_m$ as the $P_{pv}$ corresponding to the MPP. For example, if $P_{br}=25$W, the MPP is unique as 12.19W, and the corresponding unique $I_{pv}$ and $V_{pv}$ are 303.9mA and 40.11V, which are marked by the dots or triangles in Fig.~\ref{1550irradiance}.

\begin{figure}
	\centering
    \includegraphics[scale=0.6]{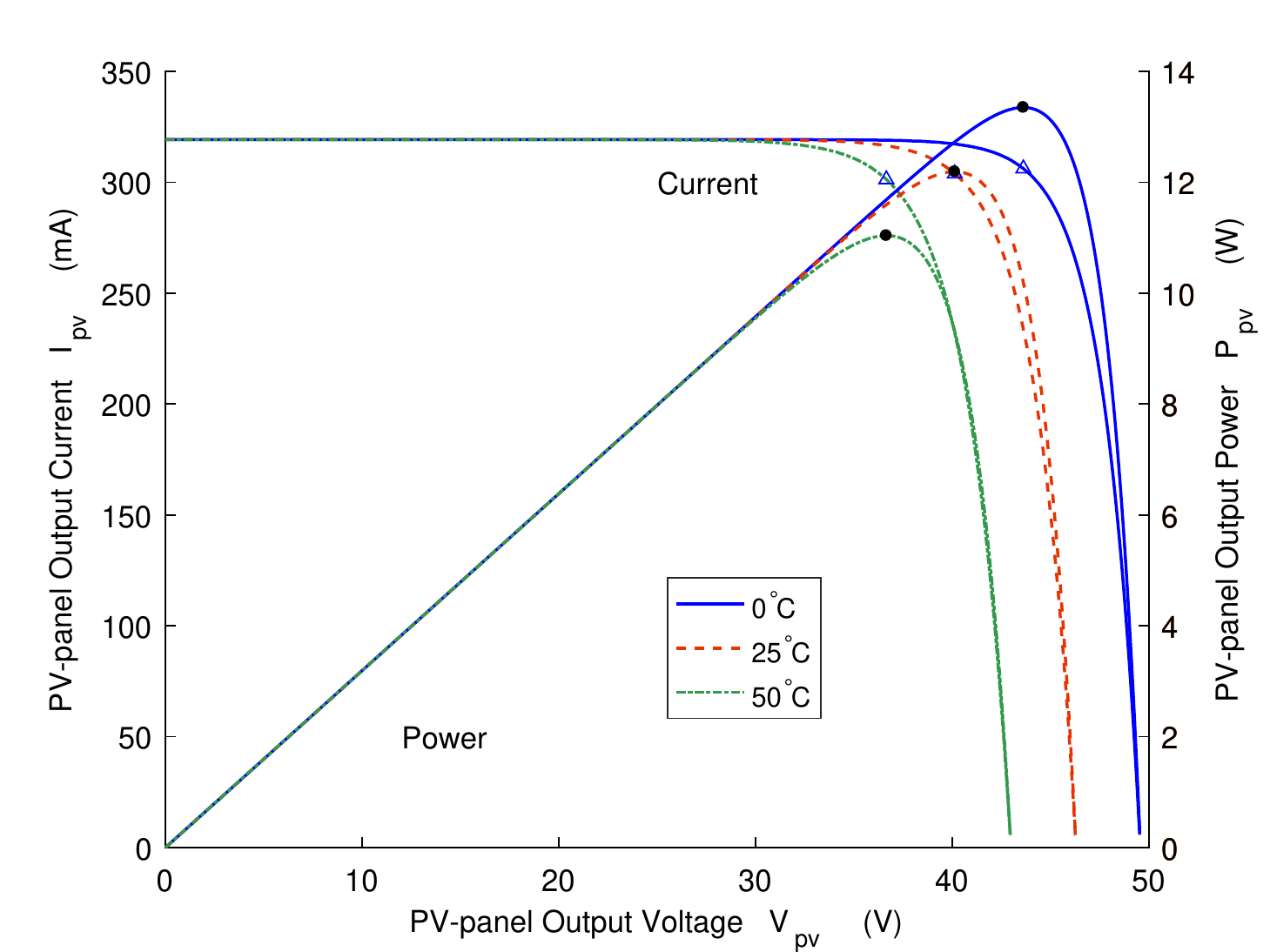}
	\caption{PV-panel Output Current and Power vs. Voltage ($P_{br} = 25$W)}
    \label{1550temperature}
\end{figure}

On the other hand, to study the effects of the PV-cell temperature $T$ on the beam-to-electricity power conversion, we carry out the simulation under three different temperatures (0$^{\circ}$C, 25$^{\circ}$C, 50$^{\circ}$C) when $P_{br}=25$W.

Fig.~\ref{1550temperature} depicts the variations of $I_{pv}$ and $P_{pv}$ on different $V_{pv}$. As can be seen, the $I_{pv}$ and $V_{pv}$ corresponding to the MPP goes down as $T$  rises. Therefore, the MPP decreases as $T$ increases.

\begin{figure}
	\centering
    \includegraphics[scale=0.6]{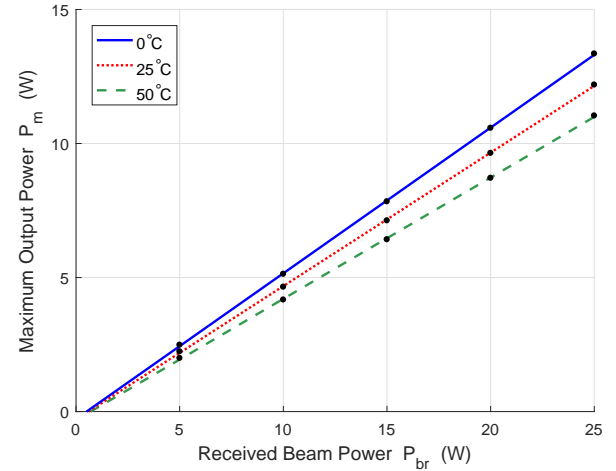}
	\caption{Maximum Output Power vs. Received Beam Power}
    \label{prpm}
\end{figure}

Thereafter, MPPs under the three temperatures when $P_{br}$ takes different values can be obtained. Dots in Fig.~\ref{prpm} show all the MPPs corresponding to different $P_{br}$. To explore the relationship between $P_m$ and $P_{br}$, we adopt the following approximation formula:
\begin{equation}\label{fprpm}
  P_{m} \approx a_2 P_{br} + b_2,
\end{equation}
where $a_2$ and $b_2$ are the curve fitting coefficients, which are listed in Table \ref{pvparamaters}. As in Fig.~\ref{prpm}, the MPP dots are all in the curves, which validates that the fitting method in \eqref{fprpm} matches the measured MPPs well.

\begin{table}[b]
\newcommand{\tabincell}[2]{\begin{tabular}{@{}#1@{}}#2\end{tabular}}
\centering
\caption{Beam Transmission Parameters}
\begin{tabular}{C{1.1cm} C{5.5cm}}
\hline
 \textbf{Parameter} &\tabincell{c}{\textbf{Visibility} \\ \ \textbf{High} \qquad\qquad \textbf{Average} \qquad\qquad \textbf{Low}} \\
\hline
\bfseries{$\beta$} & {\quad\ 3.91} \\
\bfseries{$\gamma$}   & \quad 550nm \\
\bfseries{$\nu$} & {\ 30km \qquad\qquad 11km \qquad\qquad  4km} \\
\bfseries{$\chi$}   & {\quad\quad\ 1.6 \qquad\qquad\ \ 1.3 \qquad\qquad\ \ 0.585$\nu$$^\frac{1}{3}$} \\
\hline
\label{pathloss}
\end{tabular}
\end{table}

\begin{figure}
	\centering
    \includegraphics[scale=0.6]{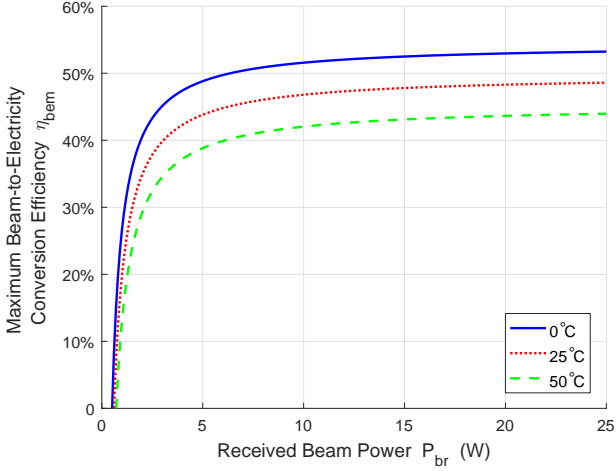}
	\caption{Maximum Beam-to-Electricity Conversion Efficiency vs. Received Beam Power}
    \label{etale}
\end{figure}

Thereafter, the maximum PV-panel conversion efficiency $\eta_{bem}$, that is the conversion efficiency when the PV-panel works at the MPPs, can be depicted as:
\begin{equation}\label{etalem}
  \eta_{bem} = \max_{P_{pv} \to P_{m}}\eta_{be}=\frac{P_{m}}{P_{br}}= a_2+\frac{b_2}{P_{br}}.
\end{equation}
Fig.~\ref{etale} shows how $\eta_{bem}$ varies with the received beam power $P_{br}$. From Fig.~\ref{etale}, $\eta_{bem}$ experiences a rapid growth at first, then it approaches to a relatively stable value. On the other hand, $\eta_{bem}$ goes down with the increment of $T$, which is consistent with how MPPs change with $T$ in Fig.~\ref{prpm}.

\section{End-to-end Performance Analysis}\label{}
In this section, we will at first investigate the relationship between the received beam power $P_{br}$ and the source power $P_s$, i.e., the end-to-end power relationship based on the above models. Then, we will analyze the maximum end-to-end power transmission efficiency. We will prove that the optimal power transmission efficiency uniquely exists. Finally, we will discuss the guidelines of system design and implementation to achieve the optimal power transmission efficiency.

Based on \eqref{plps}, \eqref{etabt} and \eqref{fprpm}, the end-to-end power relationship can be obtained as:
\begin{equation}\label{gspspm}
\begin{aligned}
  P_{m} &= a_2 \eta_{bt} P_{bt} + b_2 \\
  &= a_1 a_2 \eta_{bt} \sqrt{b_1+P_{s}} + a_2 \eta_{bt} c_1 + b_2.
\end{aligned}
\end{equation}
As can be seen, both $\eta_{bt}$ and $P_{s}$ have influences on $P_{m}$. For certain $\eta_{bt}$, how $P_{m}$ changes over $P_{s}$ can be obtained.

\begin{figure}
	\centering
    \includegraphics[scale=0.6]{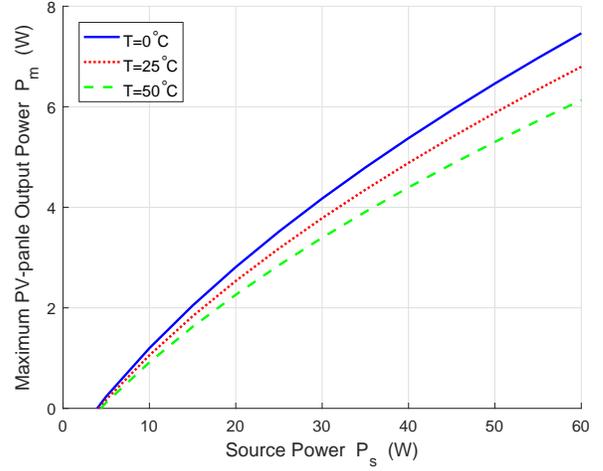}
	\caption{Maximum Output Power vs. Source Power}
    \label{1550pspm}
\end{figure}

Fig.~\ref{1550pspm} depicts the relationship when $\eta_{bt}$ takes 100\% and the PV-cell temperature $T$ takes 0$^{\circ}$C, 25$^{\circ}$C and 50$^{\circ}$C. From Fig.~\ref{1550pspm}, $P_{m}$ goes up gradually with the increment of $P_{s}$ after reaching the power threshold for generating resonant beam. At the same time, given the certain $P_{s}$, higher $P_{m}$ can be obtained when $T$ takes smaller value.

\begin{table}[b]
\newcommand{\tabincell}[2]{\begin{tabular}{@{}#1@{}}#2\end{tabular}}
\centering
\caption{Beam-to-Electricity Conversion Parameters}
\begin{tabular}{C{3.7cm} C{0.3cm} C{3.2cm}}
\hline
 \textbf{Parameter} & \textbf{Symbol} & \tabincell{c}{\textbf{Value}} \\
\hline
\bfseries{Short-circuit current}                                & {$I_{sc}$}    & \tabincell{c}{$0.305 A$}\\
\bfseries{Open-circuit voltage}                                 & {$V_{oc}$}    & \tabincell{c}{$0.464 V$} \\
\bfseries{Irradiance used for measurement}                 & {$I_{r0}$}    & \tabincell{c}{$2.7187 W/cm^2$} \\
\bfseries{Beam frequency}       & {$\gamma$}  & \tabincell{c}{$1.9355\times10^{14} Hz$} \\
\bfseries{Quality factor}                                       & {$n$}            & \tabincell{c}{$1.1$} \\
\bfseries{Number of series cells}                               & {$N$}           & $72$ \\
\bfseries{PV-panel material }                                   & { }   & \tabincell{c}{GaSb-based}  \\
\bfseries{Measurement temperature}                              & {$T$}       & \tabincell{c}{$120^{\circ}C$} \\
\bfseries{Simulation temperature}    & {}     & \tabincell{c}{$0^{\circ}C$ / $25^{\circ}C$ / $50^{\circ}C$} \\
\bfseries{$\textbf{P}_{\textbf{m}}$-$\textbf{P}_{\textbf{r}}$ curve fitting parameter}     & {$a_2$}       & \tabincell{c}{0.5434 / 0.4979 / 0.4525} \\
\bfseries{$\textbf{P}_{\textbf{m}}$-$\textbf{P}_{\textbf{r}}$ curve fitting parameter}   & {$b_2$}   & \tabincell{c}{-0.2761 / -0.2989 / -0.3209} \\
\hline
\label{pvparamaters}
\end{tabular}
\end{table}

In addition to the electricity-to-beam conversion efficiency $\eta_{eb}$, the beam transmission efficiency $\eta_{bt}$ and the maximum beam-to-electricity conversion efficiency $\eta_{bem}$, the DC-DC conversion efficiency $\eta_{dc}$ and the battery charging efficiency $\eta_{ce}$ have impacts on the maximum overall power transmission efficiency of the ARBC system $\eta_{om}$. $\eta_{dc}$ is influenced by the DC-DC converter input power $P_{m}$ and output power $P_{dc}$, which can be defined as:
\begin{equation}\label{dcequ}
\begin{aligned}
\eta_{dc} = \frac{P_{dc}}{P_{m}}. \\
\end{aligned}
\end{equation}
$\eta_{ce}$ can be expressed with $P_{dc}$ and the battery charging power as $P_{b}$, which is the ARBC system output power, as:
\begin{equation}\label{bequ}
\begin{aligned}
\eta_{ce} = \frac{P_{b}}{P_{dc}}. \\
\end{aligned}
\end{equation}

All the symbols are listed in Table \ref{convertefficiency}. 

Based on \eqref{etaeb}, \eqref{etabt}, \eqref{chic}, \eqref{etalem}, \eqref{dcequ} and \eqref{bequ}, $\eta_{om}$ can be modeled as:
\begin{equation}\label{etao}
\begin{aligned}
 \eta_{om}&=\eta_{eb} \eta_{bt} \eta_{bem} \eta_{dc} \eta_{ce}\\
                 &=\eta_{eb}\eta_{bt}(a_2+\frac{b_2}{\eta_{eb}\eta_{bt}P_{s}}) {\eta}_{dc} \eta_{ce}\\
                 &=\frac{ a_1 a_2 \eta_{bt} \sqrt{b_1+P_{s}} + (a_2 c_1 \eta_{bt}+ b_2)}{P_{s}}{\eta}_{dc} \eta_{ce}. \\
\end{aligned}
\end{equation}


From \eqref{etao}, $\eta_{om}$ relies on $P_{s}$, $\eta_{bt}$, $\eta_{dc}$, and $\eta_{ce}$. According to \cite{salem201612}, value of $\eta_{dc}$ can be 90\%. According to \cite{batteryeff},  $\eta_{ce}$ can be up to 99\%. In the following, we will illustrate how $\eta_{om}$ varies with $P_{s}$ and $\eta_{bt}$ in different scenarios.

\begin{figure}
	\centering
    \includegraphics[scale=0.6]{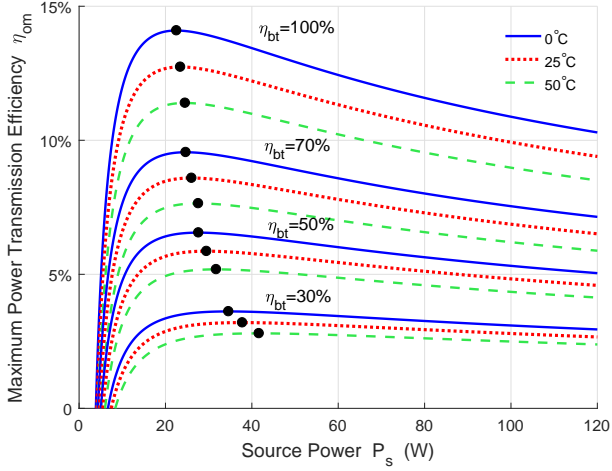}
	\caption{Maximum Power Transmission Efficiency vs. Source Power}
    \label{1550maximump}
\end{figure}

Fig.~\ref{1550maximump} shows the relationship between $\eta_{om}$ and $P_{s}$ when the beam transmission efficiency $\eta_{bt}$ varies from 100\% to 30\% and $T$ takes $0^{\circ}C$, $25^{\circ}C$ and $50^{\circ}C$. As can be seen, $\eta_{om}$ climbs up rapidly with $P_{s}$ increasing at first, then it goes down and takes a downward trend. Moreover, for certain $\eta_{bt}$, $\eta_{om}$ takes smaller value as $T$ increases.

Moreover, there seems to be a peak point of $\eta_{om}$ for each curve in Fig.~\ref{1550maximump}. To verify its existence and uniqueness when $\eta_{bt}$ takes a certain value and the variable $P_s$ lies in the interval $(0,+\infty)$, we take the derivation of $\eta_{om}$ with respect to $P_s$ as:
\begin{small}
\begin{equation}\label{detaomdps}
\begin{aligned}
&\frac{d\eta_{om}}{dP_s}\\
&=\bigg[a_2\eta_{bt}\frac{a_1P_s/(2\sqrt{b_1+P_s})-(a_1\sqrt{b_1+P_s}+c_1)}{P_s^2}-\frac{b_2}{P_s^2}\bigg]\eta_{dc} \eta_{ce}\\
                       &=\frac{a_2\eta_{bt}\eta_{dc} \eta_{ce}}{P_s^2\sqrt{b_1+P_s}}\bigg[\frac{a_1P_s}{2}-a_1(b_1+P_s)-(c_1+\frac{b_2}{a_2\eta_{bt}})\sqrt{b_1+P_s}\bigg]\\
                       &=\frac{a_2\eta_{bt}\eta_{dc} \eta_{ce}}{(t^2-b_1)^2t}\bigg[-\frac{a_1}{2}t^2-(c_1+\frac{b_2}{a_2\eta_{bt}})t-\frac{a_1b_1}{2}\bigg]\\
                       &\qquad(\text{where}\ t=\sqrt{b_1+P_s})\\
                       &:=\frac{a_2\eta_{bt}\eta_{dc} \eta_{ce}}{(t^2-b_1)^2t}g(t),
\end{aligned}
\end{equation}
\end{small}
where $g(t)=-a_1t^2/2-(c_1+b_2/(a_2\eta_{bt}))t-a_1b_1/2$ is a quadratic function. So, the sign of the derivation depends on the function $g(t)$.

We define $g(\sqrt{b_1})=-a_1b_1-(c_1+b_2/(a_2\eta_{bt}))\sqrt{b_1}$. With the help of MATLAB, we find $g(\sqrt{b_1})>0$ for all the cases. Moreover, we have $g(+\infty)=-\infty$ and $g(0)=-a_1b_1/2<0$, since $a_1$ and $b_1$ are both greater than 0 in all the considered scenarios. By the well-known intermediate value theorem, there exists one and only one point $\xi\in(\sqrt{b_1},+\infty)$ that $g(\xi)=0$. Thus, when $t\in (\sqrt{b_1}, \xi)$, $\eta_{om}$ is strictly increasing, and when $t\in(\xi,+\infty)$, $\eta_{om}$ is strictly decreasing. Therefore, $\eta_{om}$ has a unique maximum value as $P_s$ changes.

\begin{table}[b]
\centering
\caption{Transmission or Conversion Efficiency}
\begin{tabular}{C{6.0cm} C{1.3cm}}
\hline
 \textbf{Parameter} & \textbf{Symbol}  \\
\hline
\bfseries{Electricity-to-beam conversion efficiency} & {${\eta}_{eb}$} \\
\bfseries{Beam transmission efficiency} & {${\eta}_{bt}$} \\
\bfseries{Beam-to-electricity conversion efficiency} & {$\eta_{be}$} \\
\bfseries{DC-DC conversion efficiency} & {${\eta}_{dc}$} \\
\bfseries{Battery charging efficiency} & {${\eta}_{ce}$} \\
\bfseries{Maximum power transmission efficiency} & {${\eta}_{om}$} \\
\bfseries{Optimal maximum power transmission efficiency} & {${\eta}_{opt}$} \\
\hline
\label{convertefficiency}
\end{tabular}
\end{table}

\begin{figure}
	\centering
    \includegraphics[scale=0.6]{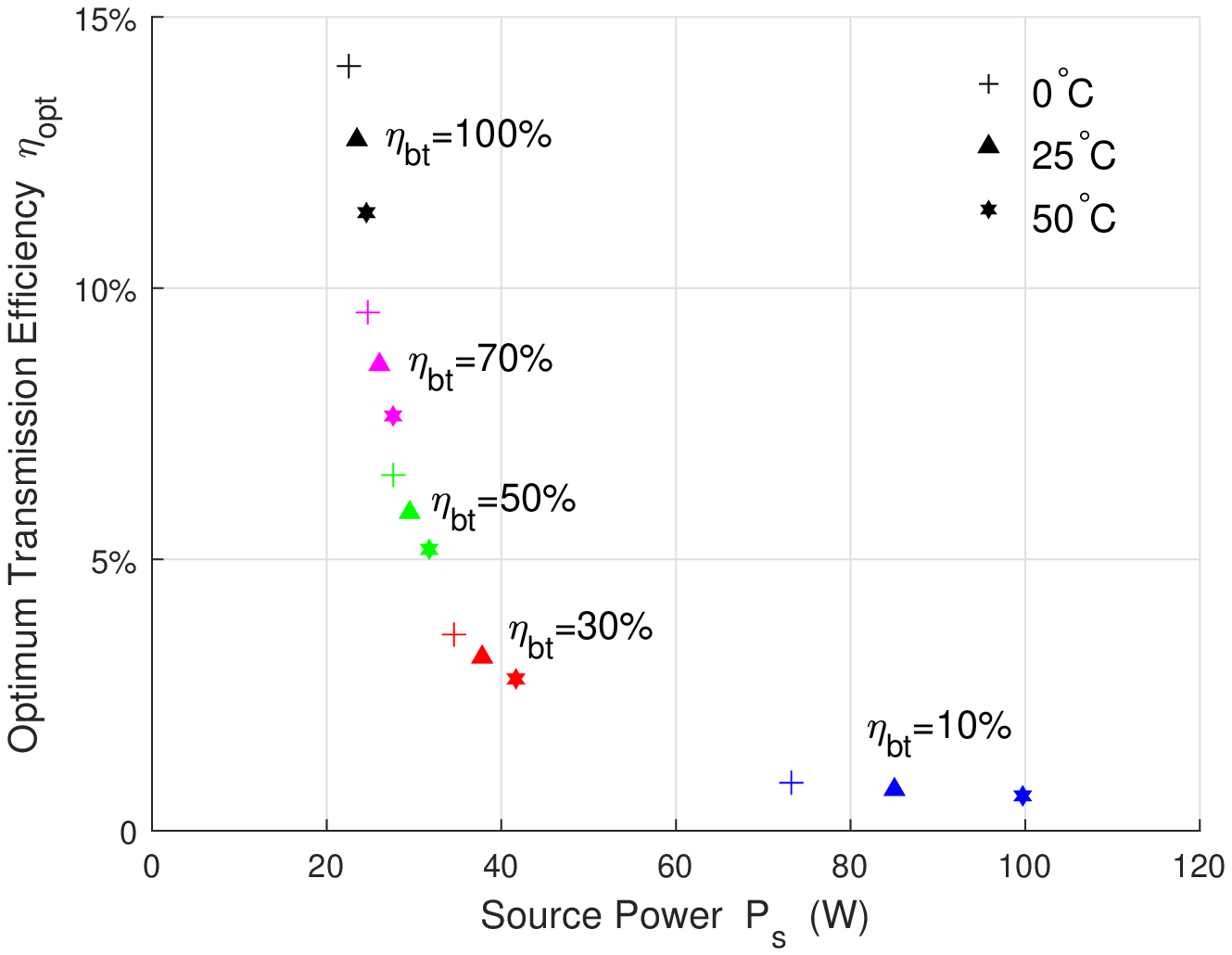}
	\caption{Optimal Transmission Efficiency vs. Source Power}
    \label{etaoptps}
%
\vspace{12pt}
	\centering
    \includegraphics[scale=0.6]{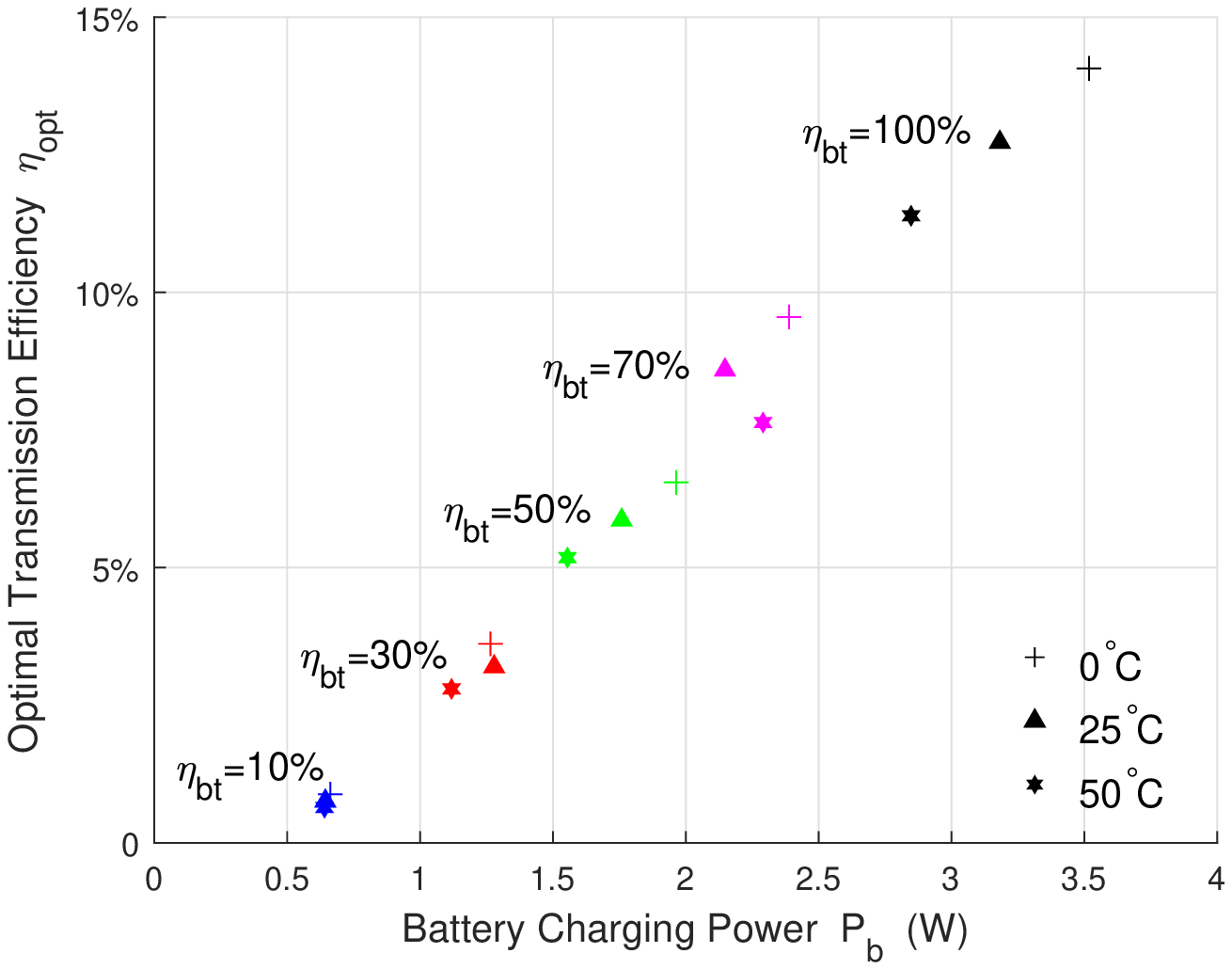}
	\caption{Optimal Transmission Efficiency vs. Battery Charging Power}
    \label{etaoptpb}
\end{figure}

Dots in Fig.~\ref{1550maximump} are the optimal points of $\eta_{om}$. To maximize the overall efficiency of the ARBC system, the system should work at the source power point corresponding to the optimal $\eta_{om}$. However, there are some cases where $P_s$ is strictly required, while the requirement for $\eta_{om}$ is not so urgent. Therefore, the unique optimal maximum power transmission efficiency point provides grounds for compromise between the source power and the overall efficiency.

We take $\eta_{opt}$ as the optimal maximum power transmission efficiency. Then, we can get how $\eta_{opt}$ changes over $P_s$ when $T$ takes 0$^{\circ}$C, 25$^{\circ}$C, 50$^{\circ}$C and $\eta_{bt}$ varies from 0 to 100\%. Fig.~\ref{etaoptps} gives the relationships. As can be seen, for certain $T$, $\eta_{opt}$ decreases dramatically from the maximum point as $P_s$ increases and $\eta_{bt}$ decreases at the beginning. Then, the decreasing trend of $\eta_{opt}$ becomes slow. If with the same $\eta_{bt}$, $\eta_{opt}$ takes lower value when $T$ increases, and $P_s$ corresponding to the $\eta_{opt}$ becomes larger. At the same time, for certain $\eta_{bt}$, $T$ has less impacts on $\eta_{opt}$ as $P_s$ goes up.

Fig.~\ref{etaoptps} provides the guidelines for source power control, so that the optimal charging efficiency can be selected to fully utilize the transmitter energy under a certain PV-cell temperature. For example, in a military application scenario, to guarantee the sensors working long enough time, batteries should be charged in time before running out. Since charging the batteries with wires is hard to implement in the military fighting environment, wireless charging becomes urgent. Sending an unmanned aerial vehicle (UAV) which carries the power source to charge the sensors can be an ideal option. To guarantee all the sensors as much energy as possible from the UAV, the UAV should charge the sensors with the optimal maximum transmission efficiency $\eta_{opt}$. Here we assume that the sensors take all the power the UAV offers regardless of the current or voltage. According to Fig.~\ref{etaoptps}, if under same $T$, $\eta_{opt}$ is determined by $\eta_{bt}$, which is influenced by the air quality and the distance between the power supplier and the sensor. Then, under the certain air quality, the UAV can control $P_s$ provided to the sensor to maximize the source power with $\eta_{opt}$.

\begin{figure}
	\centering
    \includegraphics[scale=0.6]{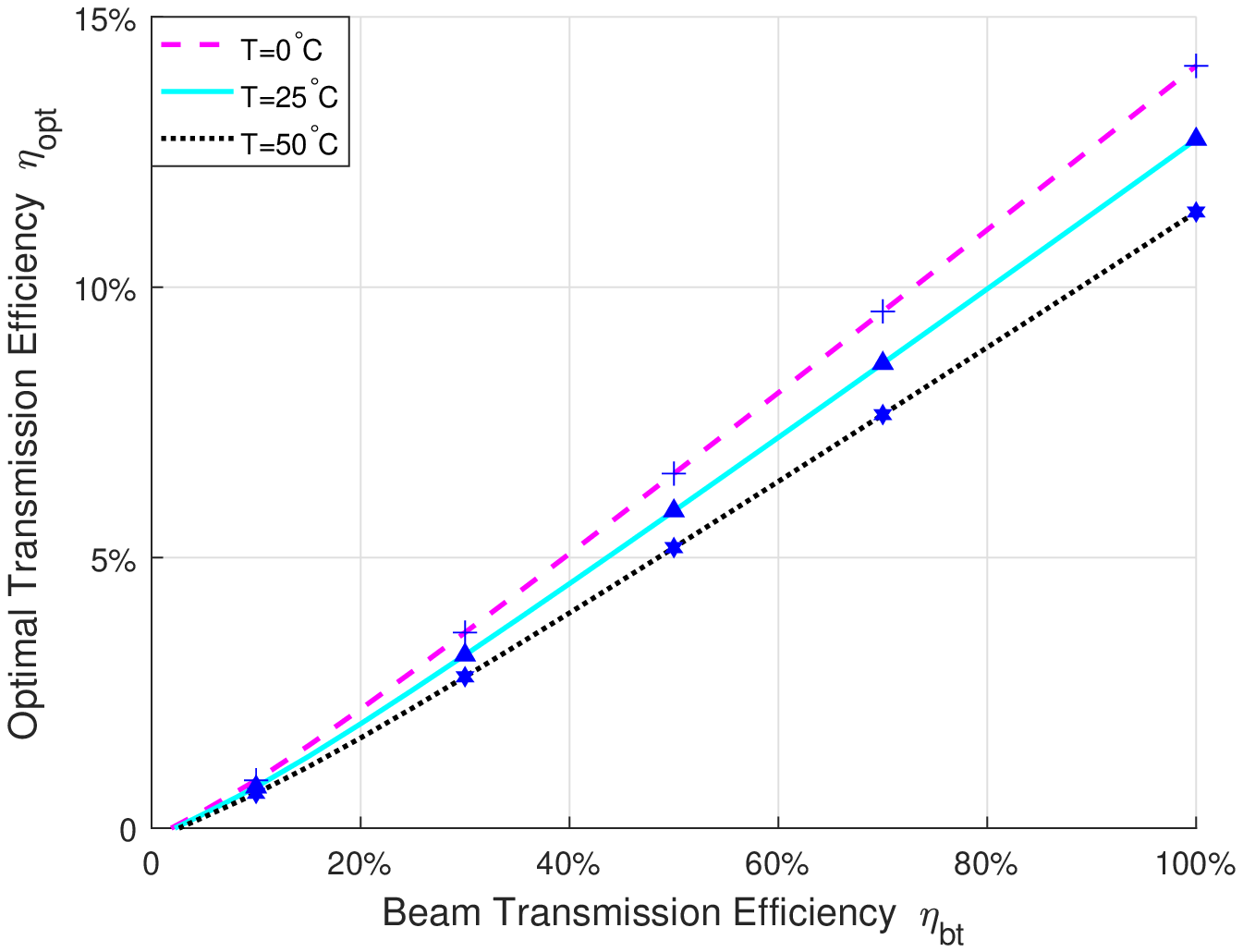}
	\caption{Optimal Transmission Efficiency vs. Beam Transmission Efficiency}
    \label{etaptetabt}
%
	\centering
    \includegraphics[scale=0.6]{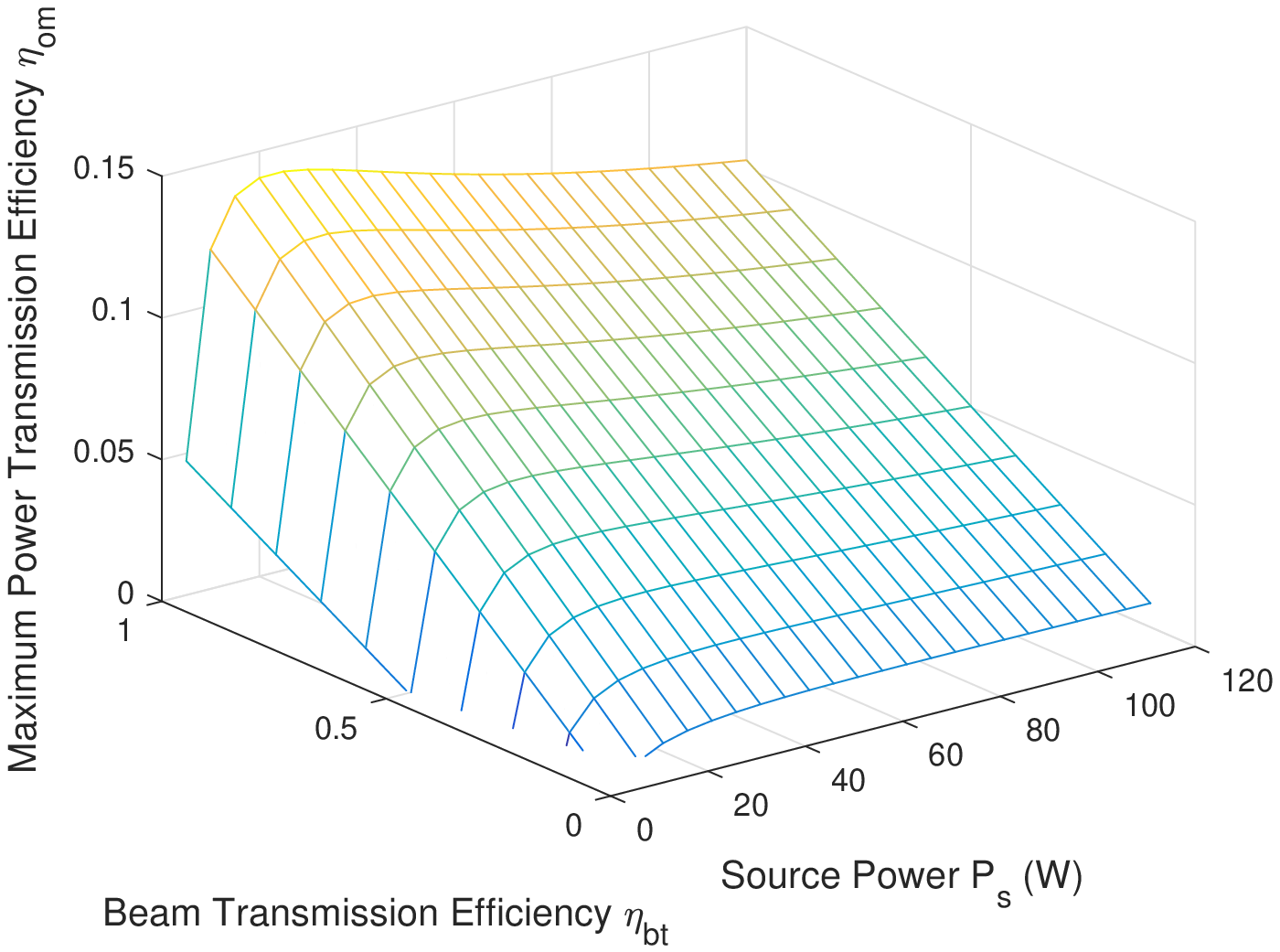}
	\caption{Maximum Power Transmission Efficiency vs. Beam Transmission Efficiency and Source Power ($T$=$0^{\circ}C)$}
    \label{etaomd}
\end{figure}

Fig.~\ref{etaoptpb} shows how the optimal transmission efficiency $\eta_{opt}$ changes over the battery charging power $P_b$. From it high $P_b$ brings high $\eta_{opt}$ with the increment of $\eta_{bt}$ under same $T$. Meanwhile, for the certain $\eta_{bt}$,  the impacts that $T$ has on $\eta_{opt}$ are much larger when $P_b$ takes bigger value. While, if with same $\eta_{bt}$, high $\eta_{opt}$ tends to be obtained at lower temperature. As is known, different vehicles have different types of batteries, and each battery has its characteristic which determines different charging power requirements. To utilize the output energy with maximum efficiency, if $\eta_{bt}$ is given, the battery type can then be determined with reference to Fig.~\ref{etaoptpb}.

From Figs.~\ref{1550maximump}, \ref{etaoptps} and \ref{etaoptpb}, the influencing factors, which include the air quality and the transmission radius, can be summarized as the beam transmission efficiency $\eta_{bt}$. To analyze the relationship between $\eta_{opt}$ and $\eta_{bt}$ at different $T$, Fig.~\ref{etaptetabt} is given. High $\eta_{bt}$ guarantees high $\eta_{opt}$. Though with slightly bend, $\eta_{opt}$ and $\eta_{bt}$ takes approximately a linear relationship. At the same time, if with same $\eta_{bt}$, $\eta_{opt}$ is much higher when $T$ takes lower value.

Therefore, the maximum power transmission efficiency $\eta_{om}$ is influenced by the source power $P_s$, the beam transmission efficiency $\eta_{bt}$, and the PV-cell temperature $T$. The relationships among the three are shown in Fig.~\ref{etaomd}. If $P_s$ and $\eta_{bt}$ are given, $\eta_{om}$ can be obtained directly. For example, when $P_s$ is 40W and the $\eta_{bt}$ is 70\%, the $\eta_{om}$ is about 9\%.

In summary, the factors influencing the overall ARBC efficiency include the source power, the beam wavelength, the air quality, the transmission radius, the PV-cell temperature, the efficiency of DC-DC and the battery charging efficiency. We obtain the following information from the above analysis:
\begin{itemize}
  \item We derive the closed-formed maximum ARBC power transmission efficiency $\eta_{om}$.
  \item We prove that the optimal value of $\eta_{om}$, i.e., the optimal end-to-end transmission efficiency $\eta_{opt}$, uniquely exists as $P_s$ changes.
  \item We analyze the relationships among $\eta_{opt}$, $P_s$, $P_b$, and $\eta_{bt}$ under different circumstances, which provide design and development guidelines for the ARBC system.
\end{itemize}


\section{Conclusions}\label{Section4}

In this paper, we present a multi-module analytical model of the adaptive resonant beam charging system for electric vehicles (EVs) in the internet of intelligent vehicles (IoIV). Based on the real measurements, the quadratic curve-fitting function is adopted for the electricity-to-beam power conversion. Thus, the closed-form formula of the end-to-end power transmission efficiency is obtained. Moreover, we prove that the optimal power transmission efficiency uniquely exists. After analyzing the optimal power transmission efficiency based on the source power, the output power, and the beam transmission efficiency, the design and implementation guidelines to optimize the end-to-end power transmission efficiency are provided.

There are some open issues to be studied for future work. For example:
\begin{itemize}
  \item The electricity-to-beam conversion study in this paper is limited by the measured data availability. Thus, investigating the expanded power range is worth to pursue in the future.
  \item To enable EVs to access to the IoIV anytime and anywhere, different battery types should be investigated. The research on the characteristics of different batteries is necessary to optimize the battery charging performance.
\end{itemize}

\bibliographystyle{IEEEtran}
\bibliographystyle{unsrt}
\bibliography{references}

\end{document}